\documentclass[aps,prd,floatfix,showpacs]{revtex4}
\usepackage[dvips]{graphicx}
\newcommand{\be}{\begin{equation}}
\newcommand{\ee}{\end{equation}}
\newcommand{\bea}{\begin{eqnarray}}
\newcommand{\eea}{\end{eqnarray}}
\newcommand{\bitem}{\begin{itemize}}
\newcommand{\eitem}{\end{itemize}}

\newcommand{\benum}{\begin{enumerate}}
\newcommand{\eenum}{\end{enumerate}}
\newcommand{\bc}{\begin{center}}
\newcommand{\ec}{\end{center}}
\begin{document}
\title{INVARIANT RELATIONSHIPS DERIVING FROM CLASSICAL SCALING TRANSFORMATIONS}
\author{Sidney Bludman}
\email{sbludman@das.uchile.cl}
\homepage{http://www.das.uchile.cl/~sbludman}
\affiliation{Departamento de Astronom\'ia, Universidad de Chile, Santiago, Chile}
\author{Dallas C. Kennedy}
\email{dkennedy@mathworks.com}
\homepage{http://home.earthlink.net/~dckennedy}
\affiliation{The MathWorks, Inc., 3 Apple Hill Drive, Natick, Massachusetts, USA}
\date{\today}
\begin{abstract}
Because scaling symmetries of the Euler-Lagrange equations are generally
not variational symmetries of the action, they do not lead to
conservation laws.  Instead, an extension of Noether's theorem reduces the equations of motion to evolutionary laws that prove useful, even if the transformations are not generalized symmetries of the equations of motion.  In the case of scaling, symmetry leads to a {\em scaling evolutionary law}, a
first-order equation in terms of scale invariants, linearly relating kinematic and dynamic degrees of freedom.

This scaling evolutionary law appears in dynamical
and in static systems. Applied to dynamical central-force systems, the scaling evolutionary equation leads to generalized virial laws, which linearly connect the kinetic and potential energies. Applied to barotropic hydrostatic spheres, the scaling evolutionary equation linearly connects the gravitational and internal energy densities.  This implies well-known properties of polytropes, describing degenerate stars and chemically homogeneous non-degenerate stellar cores.
\end{abstract}
\maketitle
\tableofcontents

\section{SCALING SYMMETRY NOT GENERALLY A SYMMETRY OF THE ACTION} 

Action principles dominate physical theories because they admit transformations among dynamical variables and
exhibit common structural analogies across different systems. If
these transformations are symmetries of the action, then by Noether's theorem, they give rise to {\em conservations
laws} that reduce the number of degrees of freedom.  This relationship between symmetries of the action (variational
symmetries) and conservation laws is central to Lagrangian dynamics.  But, even if these transformations are not symmetries of the action, they nonetheless lead to useful {\em Noether's identities}.
Although equations of motion do not require Lagrangian expression, we apply this identity to transformations that are not variational symmetries; in particular, to scaling symmetry, which is generally a generalized symmetry only of the equations of motion (Section~II).  Variational symmetries and generalized symmetries both reduce the equations of motion to first order, but in different ways:

    $\bullet$ Variational symmetries imply {\em conservation laws}, first integrals of the equations of motion.

    $\bullet$ Scaling symmetry generally implies only an {\em evolutionary equation}, which reduces the equations of motion to first order in scaling invariants.

Applied to dynamical systems of bodies interacting via inverse power-law potentials, these scaling evolutionary equations are generalized virial theorems (Section~III). Applied to self-gravitating barotropic spheres in hydrostatic
equilibrium (Section~IV), the scaling evolutionary equation will be an analogous first-order equation
between scaling invariants.  In this way,
scaling evolutionary equations will illuminate the physical consequences of scaling symmetry.

So as to focus on scaling evolutionary equations, we relegate our Lagrangian formulation of
barotropic hydrostatics to Appendix A, and needed stellar thermohydrodynamics to Appendix B.
A following paper \cite{BludKenII} will derive the well-known properties of polytropes
and of homogeneous stellar cores from Section IV of the present paper.

We do not consider applications to quantum field theories~\cite{Peskin}, involving the symmetry of the
vacuum as well as the Lagrangian, which lead to important quantum anomalies and to
topological symmetries generated by topological charges.

\section{NOETHER'S THEOREM EXTENDED TO NONINVARIANT TRANSFORMATIONS} 

\subsection{Noether's Identity Implies Either Conservation Laws or Evolutionary Equations} 

We start with a system of particles described by the Lagrangian
$\mathcal{L}(t,q_i,\dot{q_i})$ and action $S=\int{\mathcal{L}(t,q_i,\dot{q_i})} dt$, where the dot designates the
partial derivative $\partial/\partial t$ with respect to the independent variable and the Einstein summation convention is assumed.  Under an infinitesimal point transformation
$\delta (t,q_i), \delta q_j (t,q_i)$ generated by $\delta t \cdot\partial/\partial t+
\delta q_i\cdot\partial/\partial q_i$,
velocities and Lagrangian transform locally as
\bea \delta \dot{q_i}=\frac{d \delta q_i}{d t}-\dot{q_i}\frac{d \delta t}{d t} \quad , \nonumber \\
\delta \mathcal{L}=[\delta t\cdot\partial/\partial t+
\delta q_i\cdot\partial/\partial q_i+\delta\dot{q_i}\cdot\partial/\partial\dot{q_i}]\mathcal{L}=
\dot{\mathcal{L}}\delta t+(\partial\mathcal{L}/\partial q_i)\delta q_i+(\partial\mathcal{L}/\partial\dot{q_i})
\Bigl[\frac{d \delta q_i}{d t}-\dot{q_i}\frac{d \delta t}{d t}\Bigr] \quad ,\eea
where $d/dt\equiv\partial/\partial t+\dot{q_i}\cdot\partial/\partial q_i+\ddot{q_i}\cdot\partial/\partial \dot{q_i}$ is the total derivative.
The canonical momentum and energy
\be p_i(t,q_i,\dot{q_i}):=\partial\mathcal{L}/\partial\dot{q_i} \quad ,
\quad E(t,q_i,\dot{q_i}):=\dot{q_i}( \partial\mathcal{L}/\partial\dot{q_i} )-\mathcal{L} \ee
have total derivatives
\be \frac{d p_i}{d t}=\partial\mathcal{L}/\partial q_i- \mathcal{D}_i \quad ,
\quad -\frac{d E}{dt}=\dot{ \mathcal{L} }+\mathcal{D}_i\cdot \dot{q_i} ,\nonumber
\ee
in terms of the Euler-Lagrange variational derivative $\mathcal{D}_i:=
  \partial\mathcal{L}/\partial q_i-d(\partial\mathcal{L}/\partial\dot{q_i})/dt$.
Since
\be \frac{d}{dt} (p_i \delta q_i)=(\partial\mathcal{L}/\partial q_i -\mathcal{D}_i) \delta q_i +
p_i\cdot \frac{d(\delta q_i)}{dt}\quad ,
\quad -\frac{d E\delta t)}{dt}=(\dot{\mathcal{L}}+\mathcal{D}_i\cdot \dot{q_i}) \delta t -
E\cdot\frac{d(\delta t)}{dt}\quad ,\ee
the {\em Noether charge}
\be G:=\mathcal{L}\cdot\delta t+p_i\cdot(\delta q_i-\dot{q_i}\delta t)= -E \delta t+p_i\delta q_i\ee
has total derivative
\be \frac{dG}{dt}=\delta\mathcal{L}+\mathcal{L}\cdot\frac{d(\delta t)}{dt}-
\mathcal{D}_i\cdot(\delta q_i-\dot{q_i}\delta t)=\bar{\delta}\mathcal{L}-
  \mathcal{D}_i\cdot(\delta q_i-\dot{q_i}\delta t)\quad , \label{eq:Noetheridentity} \ee where
$\bar{\delta}\mathcal{L}:=\delta\mathcal{L}+\mathcal{L}\cdot (d\delta t/dt)$
is the change in Lagrangian at a fixed point.

The variation in action between fixed end points is
\be \delta S_{12}=\int_2^1 dt\ \delta\mathcal{L}=\int_2^1 dt\ \Bigl[\frac{dG}{dt}-
 \mathcal{L}\cdot\frac{d(\delta t)}{dt}+\mathcal{D}_i\cdot(\delta q_i-\dot{q_i}\delta t)\Bigr]=
  G(1)-G(2)+\int_2^1 dt\ \Bigl[\delta q_i\cdot\mathcal{D}_i+\delta t\cdot \Bigl(\frac{dh}{dt}+
  \frac{\partial\mathcal{L}}{\partial t}\Bigr) \Bigr]\quad , \ee
after integrating the term in $d(\delta t)/dt$ by parts.
The action principle asserts that this variation vanishes for independent variations $\delta q_i, \delta t$ that vanish at the end points.  It implies the Euler-Lagrange equations and $dh/dt=-\partial\mathcal{L}/\partial t$. On-shell, where
the equations of motion $\mathcal{D}_i=0$ hold,
\bea \delta S_{12}=\int_2^1{ \bar{\delta}\mathcal{L}}\ dt=G(1)-G(2) \\
\fbox{$ \displaystyle \frac{dG}{dt} =\bar{\delta}\mathcal{L}$}\quad . \eea
This is {\em Noether's Equation}, an evolutionary equation for any generator in terms of the Lagrangian transformation that it generates. It expresses the equations of motion as the time derivative of the Noether charge.

\subsection{Variational Symmetries Imply Conservation Laws} 

The most general and important applications of Noether's Equation are to variational symmetries and to generalized symmetries of the equations of motion, which both preserve the stationary action principle $\delta S_{12}=0$ but reduce the equations of motion to first order in different ways.

Variational symmetries preserve the action $\delta S_{12}=0$ because $\bar{\delta}\mathcal{L}=0$ or $dB/dt$, the total derivative of some {\em gauge term} $B(t,q)$.  Noether's Equation $(d/dt)(G-B) \equiv -
\mathcal{D}_i\cdot({\delta q_i}-\dot{q_i}\delta t)$ conserves $G - B$ on-shell, where the equations of motion hold.  This original {\em Noether's Theorem}, identifying conservation laws with space-time variational symmetries, is familiar in the symmetry of central-force systems~(\ref{eq:centralforce}) under time translations and spatial rotations, leading to conservation of energy $E$ and angular momentum $l$:
\be E:=(\dot{r}^2+r^2\dot{\theta}^2)/2+V(r)\quad , \quad l:= m r^2 \dot{\theta}\quad .\ee

An illustration of how gauge terms $B(t,\mathbf{q}_i)$ conserve $G-B$, rather than the Noether charge $G$, is
the many-body system of particles with interparticle forces that depend only on the relative separations
$\mathbf{q}_i -\mathbf{q}_j$ and relative velocities $\dot{\mathbf{q}}_i -\dot{\mathbf{q}}_j$.  This system admits the {\em infinitesimal boost transformations}
\be\delta \mathbf{q}_i = \delta\mathbf{v}\cdot t\quad ,\quad \delta t=0\quad ,\quad \delta V=0\quad ,
\quad \delta K=\delta\mathcal{L}=\mathbf{P}\cdot\delta \mathbf{v}\quad ,\ee
where $M,\mathbf{P},K$ are the total mass, momentum, and kinetic energy. Because boosts change all the momenta,
the charge $G=(\mathbf{P} \cdot \mathbf{v}) t$ is not conserved.  Instead, Noether's equation gives the conservation law $(\mathbf{P}-M\dot{\mathbf{R}}) \cdot
\delta \mathbf{v}=0$, or  $M\dot{\mathbf{R}}=\mathbf{P}$, for arbitrary infinitesimal  $\delta\mathbf{v}$. While boosts change the total momentum $\mathbf{P}$, the
center-of-mass moves with velocity $\dot{\mathbf{R}}$.  This familiar center-of-mass theorem follows directly from boost symmetry, irrespective of the internal forces. It is paradigmatic for distinguishing between the effects of internal and external forces in many-body system.

Conversely conservation laws imply invariance of the Lagrangian modulo a possible gauge term, so that
variational symmetries imply {\em conservation laws}.

\begin{table}[b] 
\caption {Period-Amplitude Relations and Virial Theorems for Inverse Power-Law Potentials $V \sim 1/r^n$}
\begin{tabular}{|l|l|l|l|}
\hline\hline
$n$&System&Period-amplitude relation $t\sim r^{1+n/2}$&Virial theorem\\
\hline\hline
-2 & isotropic harmonic oscillator             & period independent of amplitude& $\langle K\rangle =\langle V\rangle$ \\
-1 & uniform gravitational field                      & falling from rest, e.g., $z=gt^2 /2$  & $\langle K\rangle =\langle V\rangle /2$ \\
0  & free particles                                 & constant velocity $r\sim t $          & $\langle K\rangle =0$                \\
1  & Newtonian potential            & Kepler's Third Law $t^2\sim r^3$      & $\langle K\rangle =-\langle V\rangle /2$ \\
2  & inverse-cube force             & $t \sim r^2 $                         & $\langle K\rangle =-\langle V\rangle $ \\
\hline\hline
\end{tabular}
\end{table}

\subsection{Generalized Symmetries Reduce the Equations of Motion to First Order} 

The remainder of this paper will deal with {\em generalized symmetries of the equations of motion}, which are not variational symmetries and do not lead to conservation laws, but to {\em transformation equations}.
These generalized transformations, nevertheless, reduce the equations of motion to first order.
For example, a nonrelativistic particle, with radial, transverse and angular momenta
\be \mathbf{p}_r=(\mathbf{r}\cdot\mathbf{p})\mathbf{r}/r^2\quad , \quad \mathbf{p}_t=\mathbf{p}-\mathbf{p}_r =-\mathbf{r}\times\mathbf{l}/r^2 \quad , \quad \mathbf{l}:=\mathbf{r}\times \mathbf{p}\quad ,\ee
and energy
\be E (\mathbf{r},\mathbf{p})=K-V(r):=\mathbf{p}^2/2m -V(r) \quad , \quad p^2=p_r ^2+l^2/r^2
  \quad ,\label{eq:centralforce}\ee
in an inverse-power central potential $V(r) \sim 1/r^n$. The Lagrangian for this system
$\mathcal{L}(\mathbf{r},\mathbf{p})=K+V(r):=\mathbf{p}^2/2m +V(r)$
admits the {\em scaling symmetry}
\be \delta t=\beta t\quad , \quad \delta\mathbf{r}=\mathbf{r}\quad , \quad \delta\mathbf{p}=(1-\beta)\mathbf{p}\quad , \quad \delta K=2(1-\beta)K\quad ,\quad \delta V=-n V ,\ee
generated by
\be G_{nA}:=-\beta E t +\mathbf{r}\cdot\mathbf{p},\ee
provided $\beta\equiv 1+n/2$, so that the Lagrangian is homogeneous of degree $-n$, $\delta \mathcal{L}=2(1-\beta)\mathcal{L}=-n\mathcal{L}$.
Since $\delta (r ^{\beta}/t)=0$, all distances scale with time as $r_i\sim t^{1/\beta}$, as shown 
in Table I ~\cite{Landau}. 

The {\em scaling evolutionary equation}
\be dG_{nA}/dt=\delta \bar{\mathcal{L}}=(1-n/2)\mathcal{L}\quad ,\ee
linearly connects the kinetic and potential energies to the time derivative of the single particle {\em virial} $A:=\mathbf{r}\cdot\mathbf{p}=r p_r$.  This scaling evolutionary equation
is the {\em virial equation}
\be d (r p_r)/dt=(1-n/2)(K+V)+(1-n/2)(K-V)=2K+n V \quad .\ee
Only for zero-energy orbits $E=K+V=0$ in an inverse-cube force $n=2$, is the virial $A$ conserved and scaling a variational symmetry.

The first-order {\em orbit equation} for the scaling invariant
\be  r d\theta/d r=p_t/p_r=1/\sqrt{r^2 p^2/l^2-1} =1/\sqrt{2 m r^2 [E-V(r)]/l^2-1} ,\ee
can be solved by quadratures
\be \theta(r)=\theta_0 + \int_{r_0}^r dr/\bigl\lbrace r \sqrt{2 m r^2 [E-V(r)]/l^2-1}\bigr\rbrace .\ee
(In the Kepler Problem $V(r)=-k/r$, the integrals reduce to elementary functions and the bound orbits are ellipses
 \be r(\theta)=a (1-\epsilon^2)/[1-\epsilon\sin (\theta-\theta_0)] ,\ee
of eccentricity $\epsilon:=\sqrt{1-l^2/mk a}$ and semimajor axis $a:=-k/2E$.)
From this first-order equation and angular momentum conservation,
the temporal evolution is
\be dt=(m r^2/l) d\theta=m d r/\sqrt{2 m (E-V)(r))-(l/r)^2}\quad , \ee
so that
\be t(r)=t_0+\int_{r_0}^r dr/\sqrt{2 r^2 [E-V(r)]-(l/m)^2} \ee
completes the integration of the central-force problem.

\subsection{Even Transformations That Are Not Symmetries Lead to Useful Evolutionary Equations} 
For the same system, the {\em radial translation}
\be \delta t=\beta t\quad ,\quad \delta\mathbf{r}=\mathbf{r}/r\quad ,\quad \delta\mathbf{p}=-\mathbf{p}_t/r\quad ,\ee
generated by
\be G_{nB}:=\mathbf{p}\cdot\delta\mathbf{r}= p_r \ee
is not a generalized symmetry.
The {\em radial translation evolutionary equation}
\be dp_r/dt=l^2/m r^3-dV/dr \ee
is the radial equation of motion.

Generalized symmetry or not, the first-order evolutionary equations (17) for $r p_r$ and (25) for $p_r$  are useful expressions for the radial equation of motion. While the virial $A$ and radial momentum $p_r$ are not conserved,
their {\em time averages} $\langle dA/dt\rangle = \langle dp_r/dt \rangle=0$ in any bounded
ergodic system, so that, for time averages
\begin{description}
\item[scaling transformation imply] $2\langle K\rangle =-n\langle V\rangle$, the usual virial law generalized to arbitrary inverse-power potentials;
\item[radial displacement transformation imply] $(l^2/m)\langle 1/r^3\rangle =-n\langle V/r\rangle,~l \neq 0$, useful for relativistic corrections to noncircular Coulombic ($n=1$) orbits~\cite{Schwinger}.
\end{description}

The next section considers energy-conserving many-body systems
$\mathcal{L}$=$\mathcal{L}(\mathbf{r}_i,\dot{\mathbf{r}}_i)$, for which
the scaling evolutionary equation is a generalized virial law.  Section~IV, treats the hydrostatics of barotropic spheres, for which the Lagrangian $\mathcal{L}(r,H,H')$ depends explicitly on the
radial variable $r$.  Instead of a first integral, In both these examples, instead of
a conservation law, scaling symmetry implies an evolutionary equation,
a first-order differential
equation between scaling invariants~\cite{Olver,Blumen}, linearly relating the the ``kinetic'' term $K$ to the ``potential'' term $V$ in the Lagrangian $\mathcal{L}=K-V$.

\section{SCALING EVOLUTIONARY EQUATIONS} 

\subsection{Mechanical Evolutionary Equation is a Generalized Virial Law}  
The previous derivation generalizes to a nonrelativistic many-body system with particle coordinates $\mathbf{r}_i$. The scale transformation
\be \delta t=\beta\cdot t\quad , \quad \delta \mathbf{r}_i =\mathbf{r}_i\quad ,\quad \delta (\partial/\partial t)=-\beta\cdot (\partial/\partial t)\quad ,\quad \delta\dot{\mathbf{r}_i}=(1-\beta)\cdot\dot{\mathbf{r}_i}\quad ,
\quad \delta K=2(1-\beta) K
\label{eq:dynscaletrans} \ee
is generated by the Noether charge

\be G_n:=-\beta\cdot E t +A\quad , \ee
where $A:=\sum{\mathbf{p}_i \cdot \mathbf{r}_i}$ is the {\em many-body virial}.  

If the pairwise potential energies are inverse powers $V_{ij}\sim
r_{ij}^{-n}$ of the interparticle distances $r_{ij}:=|\mathbf{r}_i-\mathbf{r}_j|$, the interparticle potentials are homogeneous in their coordinates $r(dV_{ij}/dr) = -n V_{ij}, ~\delta V_{ij}=-n V_{ij}$. Scaling is still a generalized symmetry of this many-body system, provided $n\equiv 2(\beta-1)$, $\beta\equiv 1+n/2$.
Then $\delta V =-(1-\beta)\mathbf{r}\cdot \nabla V,~ \delta\dot{A}=
(1-\beta)\dot{A}$ and the total energy and Lagrangian are homogeneous functions of their arguments, scalar densities of weight $-n$,
\be   \delta E=-n E\quad ,\quad \delta\mathcal{L}=-n\mathcal{L}\quad ,\quad \bar{\delta}\mathcal{L}=\delta\mathcal{L}+\beta\mathcal{L}=(1-n/2)\mathcal{L} .\ee

Because energy is conserved, Noether's equation (9) implies the {\em scaling evolutionary equation}
\be \frac{d G_n}{d t}=-(1+n/2)\mathcal{L}+\dot{A}=
\bar{\delta}\mathcal{L}\quad ,\quad \dot{A}=(1+n/2) E+(1-n/2)\mathcal{L}=2 K+ n V\quad .\ee
This is a generalized virial law linearly
relating the time derivative of the virial $A$ to a linear combination of the nonrelativistic kinetic energy $K$ and the power-law potential $V$.

For periodic or long-time averages in bounded ergodic systems, $\langle\dot{A}\rangle$ = 0, so that the
{\em virial theorem} relates time averages $2\langle K\rangle$ = $-n\langle V\rangle $ .  Table~I tabulates these period-amplitude relations and generalized virial
theorems for orbits in the five important inverse-power-law potentials $n=-2,-1,0, 1, 2$.
Only for inverse cube forces $V(r) \sim 1/r^2$ would dynamical scaling reduce to a symmetry of the action, and the Noether charge $G_2=-2(K+V)t+A$ be conserved. For potentials more singular than $1/r^2$, there are no bound states.

\subsection{Scaling Evolutionary Equation in Classical Electrodynamics}

Noether's equation applies to continuous Lagrangian systems (fields) as well as discrete systems. In this case,
$\mathbf{r},t$ are independent variables.  If $\mathbf{f},~\mathbf{G}=
E\times\mathbf{B}/4\pi c,~\mathbf{\mathcal T},~U$
are respectively the electromagnetic force density, momentum density, momentum flux tensor, and energy density,
then momentum balance reads
\be \partial \mathbf{G}/\partial t +\nabla \cdot \mathbf{\mathcal T}+\mathbf{f} =0\quad . \ee
From this follows, the evolutionary equation :
\be \partial(\mathbf{r}\cdot \mathbf{G})/\partial t+\nabla\cdot(\mathbf{\mathcal T}\cdot\mathbf{r})-
U+\mathbf{r}\cdot\mathbf{f}=0 \quad ,\ee
an electromagnetic analogue of the mechanical virial law (31).
When time-averaged, this becomes an electromagnetic virial theorem~\cite{Schwinger}.

\subsection{Scaling Evolutionary Equation in Classical Conformal Field Theory}

In any relativistic field theory, space-time scaling (dilatation) symmetry leads to the familiar evolutionary equation
\be \frac{\partial G^{\mu}}{\partial x^{\mu}}=\Theta^{\mu}_{\mu}\quad ,\ee
where $\partial /\partial x_{\mu}$ is the four-dimensional divergence, $G^{\mu}$ is the dilatation current, and
$\Theta^{\mu}_{\mu}$ is the trace of the energy-momentum tensor~\cite{CallanColeman,Coleman,Peskin}.  If this
trace vanishes, the dilatation
charge is conserved, implying conformal symmetry.

The most familiar example of conformal symmetry is Laplace's equation in $n$ spatial dimensions. In two dimensions,
conformal symmetry implies the Cauchy-Riemann equations, so that any analytic function is a solution of
Laplace's equation.  In higher dimensions, conformal symmetry implies the conservation laws associated with
translations, rotations, dilatations, and spatial inversions.  Ignoring charges, the electromagnetic field is conformally invariant.

These familiar examples from conservative systems recall how Noether's equation leads to useful evolutionary
equations, whether or not scaling symmetry is broken. The chief purpose of this paper remains to consider the hydrostatic equilibrium of barotropic spheres, which differ by being static, not Hamiltonian
(since the radial coordinate $r$ is not ignorable), and derive from
a Least Energy variational principle, instead of a Least Action principle.

\section{SCALE-INVARIANT BAROTROPIC STARS} 

\subsection{Mechanical Structure of Barotropic Stars}

The hydrostatic structure of barotropes depends only on mass continuity and
pressure equilibrium,
\be d m/d r=4\pi r^2 \rho \quad, \quad -dP/dr=G\rho m/r^2\quad ,\quad \label{eq:masspressbalance}\ee
or the second-order equation
\be \frac{1}{r^2}\frac{d}{dr}\Bigl(\frac{r^2}{\rho}\frac{dP}{dr}\Bigr) = -4\pi G\rho\quad ,\quad
(r^2 H')'+4\pi G r^2 \rho(H)=0 \label{eq:hydrostaticEL-1}\quad ,
\ee
where $':=d/dr$. In terms of the entropy $H(r)=\int{dP/\rho}$ and specific gravitational force $d V/dr=g:=Gm/r^2$, the equation of
hydrostatic equilibrium reads
\be d (H+V)/d r=0 \quad ,\ee
so that (\ref{eq:hydrostaticEL-1}) is Poisson's equation.  We consider only barotropic stars for which
the local equation of state is $P=P(\rho)$. and the density $\rho(r)$, specific internal energy $E(r)$, specific
enthalpy $H(r)$ = $E+P/\rho$, and {\em thermal gradient} $\nabla (r):= d \log{T}/d \log{P}$ are implicit
functions of the gravitational potential $V(r)$.

This static structural equation is the Euler-Lagrange equation of the Lagrangian
\be \mathcal{L}(r,H,H')=4\pi r^2[-(H')^2/8\pi+P(H)]                                                                                                                                                                                                          ,~\label{eq:hydrostaticLagrange-2} ,\ee
derived from a minimal energy variational principle in Appendix A. It describes the radial evolution
of a static barotropic sphere.  Because the radial coordinate $r$ appears in $\mathcal{L}(r,H,H')$, the energy per radial shell $E(r,H,H')=-4\pi r^2[(H')^2/8\pi G+P(H)]$ increases while moving outwards in radius.

\subsection{In a Simple Ideal Gas, Scale Invariance Requires a Constant Entropy Gradient } 
{\em Polytropes} are barotropic spheres in which the polytropic index $n:=d\log{\rho}/d\log{H}$ or polytropic exponent $1+1/n:=d\log{P}/d\log{\rho}$ are constant (Appendix B). The pressure, specific energy, specific enthalpy, enthalpy gradient and central pressure at any point are
\be P/P_c=(\rho /\rho_c)^{1+1/n}\quad ,\quad E=n(P/\rho)\quad ,\quad H=(n+1)(P/\rho)\quad ,\quad
d\log{H}/d\log{P}=1/(n+1)\quad .\ee

The hydrostatic structure is polytropic in zero-temperature (degenerate) stars and nearly constant    in convective stars and in stars starting out on the hydrogen-burning, zero-age Main Sequence (ZAMS), where the chemical composition and energy generation are homogeneous:
\begin{description}
\item[White dwarfs and neutron stars:] nonrelativistic and extreme relativistic degenerate stars; polytropes of index $n$=3/2 and 3, respectively.
\item[ZAMS stars in convective equilibrium:] with vanishing gravithermal specific heat $C^{*}$ and uniform     entropy density.  These are $n$=3/2 polytropes.
\item[ZAMS stars in radiative equilibrium:] At zero age, our Sun was a chemically homogeneous star of mean molecular weight
    $\mu=0.61$, well-approximated by the {\em Eddington standard model} $n$=3 polytrope throughout its radiative zone, containing 99.4\% of its mass.
\end{description}

Because energy generation was extended but not uniform, our ZAMS Sun would be better approximated globally by a slightly less standard $n$=2.796 polytrope~\cite{BludKen}.  Still better nonpolytropic fits would obtain by including both nonuniform energy transport and corrections to Kramers opacity: radiative transport in the $pp$-burning lower main sequence $0.11<M/M_\odot <1.2$ gives the exponent $\xi=0.57$ in the $M-R$ relation $R\sim M^{\xi}$; convective transport in the CNO-burning upper main sequence $2 <M/M_\odot <20$ gives the exponenet $\xi=0.8$~\cite{Kippen,Hansen}.)

Because our present Sun is
chemically inhomogeneous and has convective zones, it is far from being polytropic; its best polytropic fit, with index $n=3.26$, is poor~\cite{BludKen}.

\subsection{Scaling Symmetry Implies a First-Order Equation in Scaling Invariants} 

Following Chandrasekhar~\cite{Chandra}, we define homology variables
\be u:=d \log{m}/d\log{r}=3\rho/\bar{\rho}\quad ,\quad v_n:=-d\log{(P/\rho)}/d\log{r}\quad ,
\quad w_n:=-d\log{\rho}/d\log{r}=n(r)\cdot v_n\quad , \ee
where $\bar{\rho}=3 m/4\pi r^3$ is the average mass density interior to radius $r$ and
$n(r):=d\log{\rho}/d\log{(P/\rho)}$. The central boundary condition is
\be u(0)=3\quad ,\quad v(0)=0\quad ,\quad (d v/du)_0=-5/3n\quad .\ee
The mass continuity and hydrostatic equilibrium equations~(\ref{eq:masspressbalance}) become
\be d\log{u}/d\log{r}=3-u-n(r) v\quad ,
\quad d\log{v}/d\log{r}=u-1+v-d\log{[1+n(r)]}/d\log{r}\quad ,\label{eq:autonomous} \ee
and will be autonomous only when the index $n(r)$ is constant.

In polytropes, the constant index $n$ and gradient $\nabla=1/(n+1)$ makes both these equations
\be du/d\log{r}=u(3-u-n v_n)\quad ,\quad dv_n/d\log{r}=v_n(u-1+v_n)\quad \ee
autonomous, so that they can be written as the {\em characteristic differential equations}
\be \frac{d\log{v_n}}{u-1+v_n}=\frac{d\log{u}}{3-u -n v_n}=d\log{r} \label{eq:chareqns} \ee
for the homology invariants $u,~v_n$~\footnote{
This evolutionary equation is important elsewhere in mathematical physics. In population dynamics, with $\log{r}$
replaced by
time $t$, it becomes the Lotka-Volterra equation for predator/prey evolution~\cite{Boyce,JordonSmith}. The $u v_n$
cross-terms lead to growth of the predator $v_n$ at the expense of the prey $u$, so that a population that is
exclusively prey initially ($v_n=0,u=3$) is ultimately devoured $u\rightarrow 0$. For the weakest predator/prey
interaction ($n=5$), the predator takes an infinite time to reach only the finite value $v_n=1$.
For stronger predator/prey interaction ($n<5$), the predator grows infinitely $v_n\rightarrow\infty$ in a finite time.}.

The infinitesimal scale transformation
\be \delta r=r\quad ,\quad \delta H=-\tilde{\omega_n}H\quad ,\quad \delta H'=-(1+\tilde{\omega_n}) H' \ee
is generated by the Noether charge
\be G_n:= -E\cdot r +p\cdot \delta H =-r^3[\frac{H'^2}{2}+4 \pi P(H)]-r^2\frac{H'}{G}\cdot\tilde{\omega_n}H\quad .\ee
The structure is scale-invariant if and only if $P=K\rho^{1+1/n}$, so that $H \sim P/\rho, ~P \sim H^{n+1}$ and
\be n\tilde{\omega}_n = 2+\tilde{\omega}_n\quad ,
\quad 2(1+\tilde{\omega}_n) = (n+1)\tilde{\omega}_n\quad ,\quad \tilde{\omega}_n\equiv 2/(n-1)\quad ,\quad \delta\bar{\mathcal{L}}=\tilde{\sigma}_n\mathcal{L}, \quad \tilde{\sigma}_n:=1-2 \tilde{\omega}_n=\Bigl(\frac{n-5}{n-1}\Bigr), \ee
making the gravitational potential and specific internal energy both homogeneous of degree $-2\tilde{\omega}_n$ and
\be \delta\mathcal{L}=-2\tilde{\omega_n}\mathcal{L}\quad , \quad \delta E=-2\tilde{\omega_n} E\quad .\ee
The radial derivative
\be dG_n/d r \equiv \delta\mathcal{L}+\mathcal{L}+\mathcal{D}_r\cdot \tilde{\omega_n}\cdot d(Hr)/dr\quad ,\ee
obeys the scaling evolutionary equation
\be dG_n/d r=\tilde{\sigma_n}\mathcal{L}
\label{eq:scalingnonconslaw}\quad ,\ee
connecting the gravitational and internal energy densities,
just as the virial law connected the potential and kinetic energies.

For such polytropes, we introduce dimensionless units  
\be \xi:=r/\alpha\quad ,\quad \theta_n:=H/H_c=(\rho/\rho_c)^{1/n}\quad ,\ee
and the dimensional constant
\be \alpha ^2:=\frac{(n+1)}{4\pi G}K\rho_c ^{1/n-1}=(n+1)/4\pi G \cdot (P_c/\rho_c^2)\quad ,\ee
where $\rho_c$ is the central density and
\be P_c/\rho_c:= K\rho_c ^{1/n} \quad ,\quad H_c:=(n+1)P_c/\rho_c\quad .\ee
The included mass, mass density, average mass density, and gravitational acceleration are
\bea
m=4\pi\rho_c\alpha^3\cdot(-\xi^2\theta_n ')\quad ,
\quad \rho=\rho_c\cdot\theta_n ^n\quad ,
\quad \bar{\rho}=\rho_c\cdot(-3\theta_n '/\xi)\quad ,\quad
g=4\pi\rho_c \alpha^2 (-\theta_n ')\quad .\eea
Poisson's equation~(\ref{eq:hydrostaticEL-1}), combining mass continuity and hydrostatic equilibrium, takes the dimensionless {\em Lane-Emden} form
\be \frac{d}{d\xi}\Bigl( \xi^2 \frac{d\theta_n}{d\xi}\Bigr) + \xi^2\theta_n ^n = 0\quad .\label{eq:lane-emden-1}\ee

Suppressing the subscript
$n$ on $\theta_n$ and $\theta_n ':=d \theta_n/d \xi$, the homology variables are~\cite{Chandra}
\bea u:=d\log{m}/d\log{r}=3\rho(r)/\bar{\rho}=-\xi \theta^n/\theta ' \quad ,
\quad v_n:=-d\log{(P/\rho)}/d\log{r}= -\xi\theta'/\theta \quad ,
\quad \bar{\rho}/\rho_c=-3 \theta '/\xi^2 \quad , \nonumber \\
u/v_n=\theta^{n+1}/\theta '^2\quad ,\quad uv_n=\xi^2\theta^{n-1}=(\xi^{\tilde{\omega}_n}\theta)^{n-1}\quad,
\quad (uv_n)^{1/(n-1)}=\xi^{\tilde{\omega}_n+1}(-\theta ')\quad .\eea

Extracting the dimensional constant $\mathcal{C}:=-H_c^2/G$, the Lagrangian, Hamiltonian, and Noether charge are
\bea
\mathcal{L}/\mathcal{C}=\xi^2[\frac{\theta '^2}{2}-\frac{\theta^{n+1}}{n+1}]\quad
=\xi^{-2\tilde{\omega_n}}[\frac{(uv_n^n)^{\tilde{\omega_n}}}{2}-\frac{(uv)^{1+\tilde{\omega_n}}}{n+1}]\quad ,\nonumber \\
E/\mathcal{C}          =\xi^2 [\frac{\theta '^2}{2}+\frac{\theta^{n+1}}{n+1}]\quad
=\xi^{-2\tilde{\omega_n}}[\frac{(uv_n^n)^{\tilde{\omega_n}}}{2}+\frac{(uv)^{1+\tilde{\omega_n}}}{n+1}]\quad ,\nonumber \\
G_n/\mathcal{C}=\xi^2 \Bigl[ \xi ( \frac{\theta'^2}{2}+\frac{\theta^{n+1}}{n+1})+
\tilde{\omega_n} \theta \theta' \Bigr]=\xi^{-2\tilde{\omega_n}}
\Bigl[  \xi(\frac{(uv_n^n)^{\tilde{\omega_n}}}{2}-\frac{(uv)^{1+\tilde{\omega_n}}}{n+1})+
\tilde{\omega_n} u^{\tilde{\omega_n}} v_n ^{1+\tilde{\omega_n}} \Bigr]\quad .\eea

For $n=5$, scaling symmetry is a variational symmetry and the Noether charge
\be G_5=\frac{\mathcal{C}}{2}\cdot \Bigl[(uv_5 ^5)^{3/2}-\frac{(uv_5)^{3/2}}{3} +(uv_5 ^3)^{1/2}\Bigr] \ee
is conserved.  Otherwise, the Noether charge evolves according to ~(\ref{eq:scalingnonconslaw})
\be \frac{d}{d\xi} \Bigl\{ \xi^2 \cdot \Bigl[ \xi\Bigl(\frac{\theta'^2}{2}+\frac{\theta^{n+1}}{n+1}\Bigr)+
  (\frac{2}{n-1})\theta \theta' \Bigr] \Bigr\}=\Bigl(\frac{n-5}{n-1}\Bigr)\cdot\xi^2
  \Bigl( \frac{\theta'^2}{2}-\frac{\theta^{n+1}}{n+1}\Bigr) ,\ee
which reduces to the Lane-Emden equation~(\ref{eq:lane-emden-1}). This evolutionary equation describes the
growing ratio between local
internal and (negative) gravitational energy densities
\be \frac{\theta^{n+1}/(n+1)}{\theta'^2 /2} = \frac{2}{n+1} \frac{u}{v_n}\quad ,\label{eq:energyratio}\ee
as the local energy density changes from entirely internal at the center, to entirely gravitational at
the stellar surface.

\section{CONCLUSIONS} 

We have extended Noether's Equation connecting variational symmetries to conservation laws to generalized  transformations of the Euler-Lagrange equations. The resultant evolutionary equations are useful, even when the transformations are not generalized symmetries.  But, when they are generalized symmetries, they reduce the Euler-Lagrange equations to first-order equations between invariants.

For scaling symmetries, the evolutionary equation takes a special form connecting linearly the kinematic and dynamic parts of the Lagrangian.
For nonrelativistic systems with inverse-power law potentials, the scaling evolutionary equation is a generalized virial law, linearly relating the kinetic and potential energies.  For hydrostatic systems obeying barotropic equations of state, the scaling analogous evolutionary equation is a linear relation between the local gravitational and internal energies.  From this scaling evolutionary equation,  in the following article~\cite{BludKenII}, we derive all the properties of polytropes.

\appendix
\section{LAGRANGIAN FORMULATION OF BAROTROPIC HYDROSTATICS}  

Stellar structure generally depends on coupled equations for pressure equilibrium and heat transport.  Only if the heat transport leads to a local {\em barotropic} relation $P=P(\rho)$ can the hydrostatic equations be considered
independently, without reference to the thermal structure.

\subsection{Mass Continuity and Hydrostatic Equilibrium}

In a self-gravitating isolated system in local thermodynamic equilibrium, a barotrope held at zero external
pressure has thermodynamic potential energy, the
work needed to adiabatically extract unit mass, or                                                              {\em specific enthalpy} $H(\rho)=E+P/\rho$.  Barotropic energy conservation, $d H:=d P/\rho$, makes the specific enthalpy a more natural state variable than the specific internal energy $E$, pressure $P$, or density $\rho$. The equation of hydrostatic equilibrium $-d P/dr=Gm\rho /r^2:=\rho g$ is then
\be -d H/dr=dV/dr = g = Gm(r)/r^2 \quad , \label{eq:hydrostaticeq-1}\ee
describing how this local specific enthalpy or extraction energy $H(r)$ depends on the local gravitational potential $V(r)$. Integrating, we have the energy conservation equation
\be H(r)+V(r)=-GM/R\quad ,\quad r<R \quad ,\ee
where the zeros of the gravitational potential and specific enthalpy have been chosen at infinity and at the spherical surface, respectively.

Because the gravitational potential obeys Poisson's equation
\be \nabla^2 V=\frac{1}{r^2}\frac{d}{dr}\Bigl( r^2\frac{dV}{dr}\Bigr)=4\pi G \rho \quad ,\ee
the specific enthalpy obeys the second-order equation
\be \fbox{$ \displaystyle \nabla ^2 H+4\pi G\rho (H) =0 \label{eq:hydrostaticEL-2}$} \quad .\ee

To implement the equation of hydrostatic equilibrium, we need a local barotropic relation $P(\rho)$, or $P(H)$,
$\rho (H)$, which is determined by the thermal stratification
of the static matter distribution in local thermodynamic equilibrium, and by a central boundary (regularity)
condition $(d P/d r)_0=0=(d \rho/d r)_0$. Near the origin,
\bea \rho\approx\rho_c (1- B r^2)\quad , \quad m(r)\approx\frac{4 \pi \rho_c r^3}{3}\Bigl( 1-\frac{3}{5}B r^2\Bigr) \quad ,\nonumber \\
\bar{\rho}(r):= 3 m(r)/4 \pi r^3 \approx\rho_c \Bigl( 1-\frac{3}{5} B r^2\Bigr) \approx \rho_c^{2/5} \rho^{3/5}\quad.
\eea
In terms of the homology variables $w:=-d \log{\rho}/d\log{r},~u:=d\log{m}/d\log{r}$
for the mass density and included mass, $dw/du \approx -5/3$.

\subsection{A Constrained Minimum Energy Principle for Hydrostatic Equilibrium}

In a static, self-gravitating sphere of mass $M$ and radius $R$, the {\em Gibbs free energy}
\be W:=E-TS+PV=\Omega+U\quad , \ee
in terms of the gravitational and internal energies
\be \Omega=-\int_0^M (Gm/r) dm\quad , \quad U=-\int_0^R P d\mathcal{V}\quad , \ee
where $\rho$, $E$, and $-Gm(r)/r$ are the mass density, specific internal energy, and gravitational potential,
respectively. In the Eulerian description, the radial coordinate is $r$, the enclosed volume is $\mathcal{V}=4\pi r^3/3$, and the enclosed mass $m(r)$ is constrained by mass continuity $d m(r)=\rho d \mathcal{V}$.
The Gibbs free energy
\be W = \int_0^R \mathcal{L}(r,m,m') dr = -\int_0^R 4\pi r^2[Gm\rho /r+P(\rho)] dr\quad ,\quad ':=d /dr\quad ,\ee
is the available work when expanding the sphere adiabatically at fixed external pressure. The Lagrangian $\mathcal{L}$ is the Gibbs free energy per radial shell $dr$.

The constrained minimum energy variational principle~\cite{Hansen,Chui} for hydrostatic equilibrium is that
the Gibbs free energy
be stationary ($\delta W=0$) under adiabatic deformations in specific volume $\delta \mathcal{V}_{\rho}=
d(4\pi r^2\delta r)/dm$ that vanish on the boundaries and satisfy the mass continuity constraint $m' = 4\pi r^2\rho$.
This minimum energy principle has the equation of hydrostatic equilibrium
\be \mathcal{D}_r:=Gm/r^2 + H'=Gm/r^2+dP/\rho dr = 0\quad ,\ee
as its Euler-Lagrange equation, with mass continuity as a constraint.  This equation is scale invariant if the specific enthalpy $H$ scales as $m'$.

\subsection{An Unconstrained Variational Principle} 
Using Poisson's equation to incorporate the mass continuity constraint, the gravitational energy is
\be \Omega=-\int_0 ^R (V'^2/2) 4 \pi r^2 dr \quad ,\ee
so that the second-order Lagrangian (used in Section~IV)
\be \mathcal{L}(r,H,H')=4\pi r^2[-H'^2/8\pi G+P(H) ]\ee
is unconstrained and has Euler-Lagrange equation~(\ref{eq:hydrostaticEL-2}).  The canonical momentum and Hamiltonian are
\be p:=\partial\mathcal{L}/\partial H'=-r^2 H'/G = -m\quad ,\quad
\mathcal{H}(r,H,p)=-Gp^2/2 r^2-4\pi r^2 P(H)\quad , \label{eq:hydrostaticHamilton} \ee
and the canonical equations are
\be \partial\mathcal{H}/\partial p=H'=-Gp/r^2\quad ,
\quad \partial\mathcal{H}/\partial H=-p '=m'=4\pi r^2\rho \quad .\ee
Spherical geometry makes the system non-autonomous, so that $dH/d r=-\partial\mathcal{L}/\partial r
=-2\mathcal{L}/r$ and $m'(r)$
vanish only at large $r$, when the geometry approaches being planar.

\section{STELLAR THERMOHYDRODYNAMICS} 

The structure of luminous stars depends upon the coupling between hydrostatic and thermal structures through an equation
of state $P=P(\rho, T,\mu)$, which generally depends on the local temperature and chemical composition.  But, ignoring evolution, the matter entropy is locally conserved, so that steady-state stars are in local thermodynamic equilibrium (LTE). In a fluid held in pressure equilibrium at constant external
temperature, the specific Gibbs free energy $H-TS=-V(r)$ is a minimum.
In hydrostatic equilibrium, the density $\rho(r)$, specific
internal energy $E(r)$, specific enthalpy $H(r)=E+P/\rho$, specific entropy and thermal gradient
$\nabla (r):=d \log{T}/d \log{P}$ depend implicitly on the gravitational potential $V(r)$.

In the Second Law of Thermodynamics
\be T dS= d Q=dE+P d(1/\rho)\quad ,\ee
$E,~\rho$ can be written as functions of temperature and pressure.  Clever use of thermodynamic identities then leads to~\cite{Hansen,Kippen}
\be TdS=C^{*} dT\quad ,\quad d S=c_P (\nabla-\nabla_{\rm ad}) d\log{P}\quad ,\ee
where the {\em gravithermal specific heat} $C^{*}:=d S/d\log{T}=P(1-\nabla_{\rm ad}/\nabla )$
depends on the specific heat $c_P$ and the {\em adiabatic gradient} $\nabla_{\rm ad}:=
(\partial \log{T}/\partial \log{P})_S$.
This expression
relates the local thermal gradient $\nabla (r)$ to the local {\em entropy gradient} $dS(r)/d\log{P}=c_P (\nabla-\nabla_{\rm ad})$,
which generally contains both gas and radiation entropies and varies in stars in radiative equilibrium.

\begin{description}
\item[In a simple ideal gas], the equation of state, specific internal energy, specific enthalpy and adiabatic exponent are
\be P/\rho=\frac{\mathcal{R}}{\mu} T\quad ,\quad E=c_V T\quad ,\quad H=c_P T\quad ,
\quad (d\log{P}/d\log{\rho})_S =c_P/c_V :=\gamma\quad , \ee
where $\mathcal{R}$ is the universal gas constant, $\mu$ is the molecular weight, and $c_V,~c_P=c_V+\mathcal{R}/\mu$ are the specific heats at constant volume and at constant pressure.  From the Second Law of Thermodynamics (B2), the specific entropy and thermal gradient of a simple ideal gas are
\be ~dS=c_V d\log{P}-c_P d\log{\rho}\quad ,\quad S=c_V\log{\Bigl( P/\rho^{\gamma} \Bigr)}\quad ,
\quad \nabla=d\log{H}/d\log{P}=(\gamma-1)/\gamma \quad .\ee
\item[In an ideal gas supported by {\em both} gas pressure $P_{\rm gas}=\mathcal{R}\rho T/\mu$ and radiation pressure $P_{\rm rad}=a T^4 /3$], the total pressure is
\be P=P_{\rm gas}+P_{\rm rad}=\beta P+(1-\beta)P=(\mathcal{R}\rho T/\mu)[1+(1-\beta)/\beta], \ee
where the pressure ratio
\be P_{\rm rad}/P_{\rm gas}:=(1-\beta)/\beta=(4 \mu/\mathcal{R})\cdot S_{\rm rad}\quad , \ee
is proportional to the specific radiation entropy $S_{\rm rad}=4aT^3/3\rho$.
The total specific entropy
\be S=S_g + S_{\rm rad}=(\mathcal{R}/\mu)\cdot [\log{(T^{3/2}/\rho)}+4 (1-\beta)/\beta]\quad .
\ee
is therefore constant for
a cool monatomic ideal gas ($\beta=1$) or for a radiation-dominated
(supermassive) star ($\beta \approx 0$).
\item[Bound in a polytrope of index $n$], an ideal gas has constant thermal gradient, gravithermal
specific heat, and entropy-pressure gradient:
\be \nabla=1/(n+1)\quad ,\quad  c^{*}=c_P(1-\nabla_{\rm ad}/\nabla)\quad , dS/d\log{P}=c_P(\nabla-\nabla_{ad}) \leq 0 . \ee
Except for this, a polytrope's thermal structure is unconstrained and still depends on the unspecified heat transport mechanism.
\end{description}
According to Schwarzschild's minimal entropy production criterion, convective stability requires that a star's specific entropy stay constant in convective equilibrium and increase radially outwards in
radiative equilibrium. This makes barotropic stars of mass $M$ extremal in two respects:
the central pressure is minimal in for a given radius $R$; the central pressure and
temperature                                                                                                      are maximal for a given central density. Because stellar evolution is driven by developments
in the core, these bounds drive stars toward uniform entropy in late stages of evolution~\cite{Kovetz}.

\begin{acknowledgments}
SAB acknowledges support from the Millennium Center for Supernova Science through grant P06-045-F funded
by Programa Bicentenario de Ciencia y Tecnolog\'ia de CONICYT and Programa Iniciativa Cient\'ifica Milenio de
MIDEPLAN.
\end{acknowledgments}

\bibliography{bibliographyLE} 
\end{document}